# Room-temperature bipolar valleytronic transistor in MoS$_2$/WSe$_2$ heterostructures

Chongyun Jiang[1,2,†], Abdullah Rasmita[1,†], Hui Ma[1,3,†], Qinghai Tan[1], Zumeng Huang[1], Shen Lai[1], Sheng Liu[1], Xue Liu[1], Qihua Xiong[4,5,*] & Wei-bo Gao[1,6,*]

[1]*Division of Physics and Applied Physics, School of Physical and Mathematical Sciences, Nanyang Technological University, Singapore 637371, Singapore*

[2]*College of Electronic Information and Optical Engineering, Nankai University, Tianjin 300350, China*

[3]*School of Physical Science and Technology, Tiangong University, Tianjin 300387, China*

[4]*State Key Laboratory of Low-Dimensional Quantum Physics and Department of Physics, Tsinghua University, Beijing, 100084, P.R. China*

[5]*Beijing Academy of Quantum Information Sciences, Beijing 100193, P.R. China*

[6]*The Photonics Institute and Centre for Disruptive Photonic Technologies, Nanyang Technological University, 637371 Singapore, Singapore*

*†These authors contributed equally*

## Abstract

**Valley degree of freedom in the 2D semiconductor is a promising platform for the next generation optoelectronics[1-3]. Electrons in different valleys can have opposite Berry curvature, leading to the valley Hall effect (VHE)[3-7]. However, VHE without the plasmonic structure's assistance has only been reported in cryogenic temperature, limiting its practical application. Here, we report the observation of VHE at room temperature in the MoS$_2$/WSe$_2$ heterostructures. We also uncover that both the magnitude and the polarity of the VHE in the 2D heterostructure is gate tunable. We**

**attribute this to the opposite VHE contribution from the electron and hole in different layers. These results indicate the bipolar transport nature of our valleytronic transistor. Utilizing this gate tunability, we demonstrate a bipolar valleytronic transistor. Our results can be used to improve the ON/OFF ratio of the valleytronic transistor and to realize more versatile valleytronics logic circuits.**

# Main

Valley (i.e., the carrier energy band local minima in the momentum space) offers an additional degree of freedom, which can be utilized in novel optoelectronic devices[1-3,8,9]. In the monolayer transition metal dichalcogenide (TMD) $MX_2$ (M=Mo, W and X=S, Se), the K and K' valley electrons (holes) have an opposite out-of-plane Berry curvatures[1-3]. This causes the intrinsic valley Hall effect (VHE), where the anomalous velocity (i.e., the in-plane velocity component perpendicular to the in-plane electric field) in one valley is opposite to the one in the other valley, resulting in a detectable Hall voltage[1,7,10]. Moreover, due to the opposite valley magnetic moment in the K and K' valley, these two valleys are optically addressable using circularly polarized light[3,11-16]. While VHE has been demonstrated previously, these experiments require either cryogenic temperature[4,5,17,18] or plasmonic structures[9]. These limit the practical application of VHE. Moreover, like in a conventional transistor, it is important to optimize the ON/OFF ratio in a valleytronic transistor[9].

Here, we report the experimental demonstration of the VHE in $MoS_2$/$WSe_2$ van der Waals heterostructure at room temperature. In contrast with the $MoS_2$ valleytronic transistor case, where only the VHE magnitude is gate tunable[4,9], both the magnitude and the polarity of the VHE in our valleytronic transistor are gate tunable. We attribute this full gate tunability to the opposite VHE contribution from the electron and hole in different layers, which shows the bipolar transport characteristic of our device. Such full gate tunability can be used to improve the ON/OFF ratio of the valleytronic transistor beyond the current limit and to realize more compact valleytronics logic circuits.

## VHE at room temperature

The VHE in the 2D MoS$_2$/WSe$_2$ is illustrated in Fig.1a. The in-plane electric field $\vec{E}$ induces opposite anomalous velocity for the carrier in K and K' valley[3,4]. Under circularly polarized light with energy close to WSe$_2$ exciton energy, the WSe$_2$ K (K') valley electron and valley hole can be reliably excited[15]. Due to the MoS$_2$/WSe$_2$ type-II band alignment[19-21], the holes will stay in the WSe$_2$ layer while electrons undergo charge transfer to the MoS$_2$ layer[22,23]. This electron-hole layer separation suppresses the intervalley scattering[24-26] and reduces the exciton binding energy[27,28]; both are beneficial for realizing VHE at a higher temperature. Due to the weak interlayer band hybridization in K and K' valleys[28,29], the two layers can be treated as having two independent Berry curvatures in these valleys. Moreover, due to the electron-hole layer separation, the electron and hole can have independent transport mechanisms. The ability to control the layer asymmetry and carrier transport by electrical gating can result in a gate tunable VHE polarity[5].

Figure 1b shows the optical image of the MoS$_2$/WSe$_2$ heterostructure (Sample 1). The heterobilayer has an AA-stacking alignment, determined through second harmonic generation measurement (see Fig. S1a and Supplementary Note I in the Supplementary Information). The Hall bar structure is fabricated by e-beam lithography followed by thermal evaporation and lift-off process. The sample has a back-gate electrode for studying the gate dependence of VHE. The experimental setup for the Hall voltage measurement is shown in Fig. 1c. A set of Galvo mirrors is used to scan the laser beam across the sample for optical excitation under normal incidence. The light polarization is modulated between left and right circularly polarized light with a combination of a polarizer and a liquid crystal retarder (LCR) driven by a function generator. A bias voltage $V_x$ (electrode 5 and 3) and back-gate voltage $V_g$ can be applied to the sample. The Hall voltage is detected in the direction perpendicular to $V_x$ (electrode 1 and 4). It is amplified firstly with a differential preamplifier and then with a lock-in amplifier (locked to the function generator frequency) to filter out the background noise, resulted in $V_H$ (i.e., the difference between the Hall voltage under two orthogonal polarization).

With the laser focusing on the sample, we first collect the back-reflected light while scanning the sample, resulting in the image shown in Fig. 1d. Based on the reflection scanning result, the beam size can be estimated to be in the sub-micrometer range, close to the diffraction limit. Next, we do the optically excited Hall measurement. The excitation wavelength is 726 nm, on resonance with the intralayer $WSe_2$ neutral exciton transition (see Fig. S2 in the Supplementary Information for the photoluminescence (PL) spectrum). The observation of VHE at room temperature is shown in Fig. 1e. As can be seen, $V_H$ shows a linear dependence on $V_x$ for circular polarization modulation (R-L and L-R in Fig. 1e). For linear polarization modulation (H-V in Fig. 1e), the value of $V_H$ is negligible. A similar result is obtained in another AA-stacked $MoS_2/WSe_2$ sample (Sample 2, see Fig. S3 in the Supplementary Information). In contrast, no VHE at room temperature is observed for the monolayer $MoS_2$ sample (Fig. S4 in the Supplementary Information).

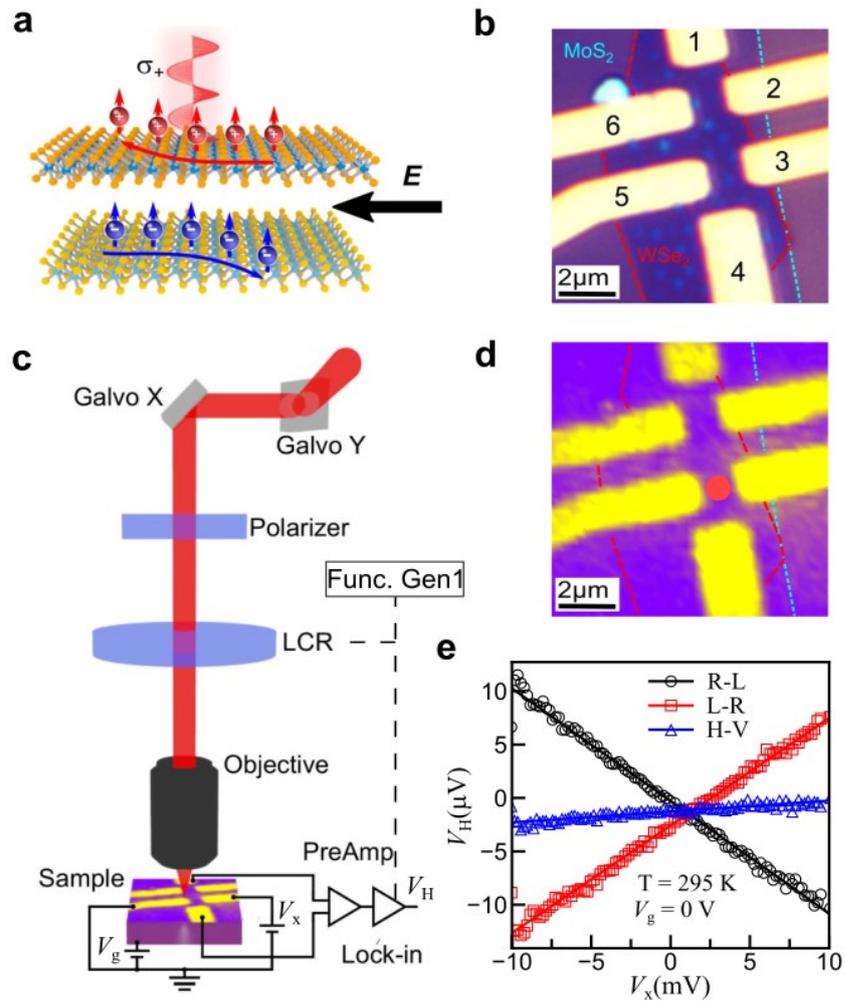

**Figure 1. VHE at room temperature.** (**a**) Illustration of VHE in MoS$_2$/WSe$_2$ heterostructure. In-plane electric field $\vec{E}$ and the carrier layer separation induced valley and layer-dependent anomalous velocity. (**b**) Optical image of the MoS$_2$/WSe$_2$ sample. The electrode numbering is for easy referencing. (**c**) Hall measurement experimental setup. The input resistance of the differential preamplifier (A-M Systems, Inc. Model 3000) is 1000 TΩ, much greater than the sample resistance. (**d**) Back-reflection image of the sample. The red dot is the excitation location for gate dependence measurement. (**e**) Observation of VHE at room temperature. The black, red, and blue symbols are the measured difference between the Hall voltage under the right and left circular polarization (R-L), under the left and right circular polarization (L-R), and under the horizontal and vertical polarization (H-V), respectively. The lines are linear fitting.

## Gate dependence measurement of VHE

To study the gate dependence of the VHE, the value of $V_H$ (right minus left circular polarization) is measured as $V_g$ is swept from -32 V to 20 V at different $V_x$ values ranging from -50 mV to 50 mV (Fig. 2a). The 220 µW, 726 nm excitation location is indicated as a red dot in Fig. 1d. Two line-cuts in Fig. 2a at $V_g = -23.5$ V and $V_g = -28.25$ V are shown in Fig. 2b. The data are fitted with the function $V_H = \alpha_H V_x + \beta_H$. The $\alpha_H$ and $\beta_H$ are proportional to Hall conductivity and circular photocurrent (CPC), respectively. As can be seen from Fig. 2b, the sign of $\alpha_H$ is changed, indicating the change in the VHE polarity. The plots of $\alpha_H$ and $\beta_H$ as a function of $V_g$ are shown in Fig. 2c. It shows that both the polarity and magnitude of the VHE are gate tunable, in contrast with the VHE in MoS$_2$ monolayer where only its magnitude is gate tunable[4,9] (see Fig. S5 in the Supplementary Information for the gate dependence of VHE in a MoS$_2$ monolayer sample). We also observed that the VHE and CPC are zero and optimum (i.e., maximum or minimum) at the same gate voltage, indicating that they share similar photovoltaic mechanisms.

Additionally, the current passing through the voltage source $V_x$ (denoted as $I_x$) is also measured. The inset of Fig. 2d shows the plot of $I_x$ versus $V_x$ at three different gate voltages showing a linear behavior with a zero offset. The longitudinal conductivity ($\sigma_{xx}$) gate dependence, obtained from the slope of $I_x$ versus $V_x$, is shown in Fig. 2d. The conductivity within the considered gate voltage range is > 0.01 µS, corresponding to a sample resistance of < 100 MΩ (see the log-linear plot in Fig. S6 in the Supplementary Information). The conductivity is monotonically increasing with the gate voltage that the sample stays in one doping regime (n-doped). By comparing Fig. 2c and Fig. 2d, we can conclude that the VHE and CPC become significant near the neutral doped regime. In this regime, both photogenerated electron and hole can induce the Hall voltage and photocurrent, as we discussed below.

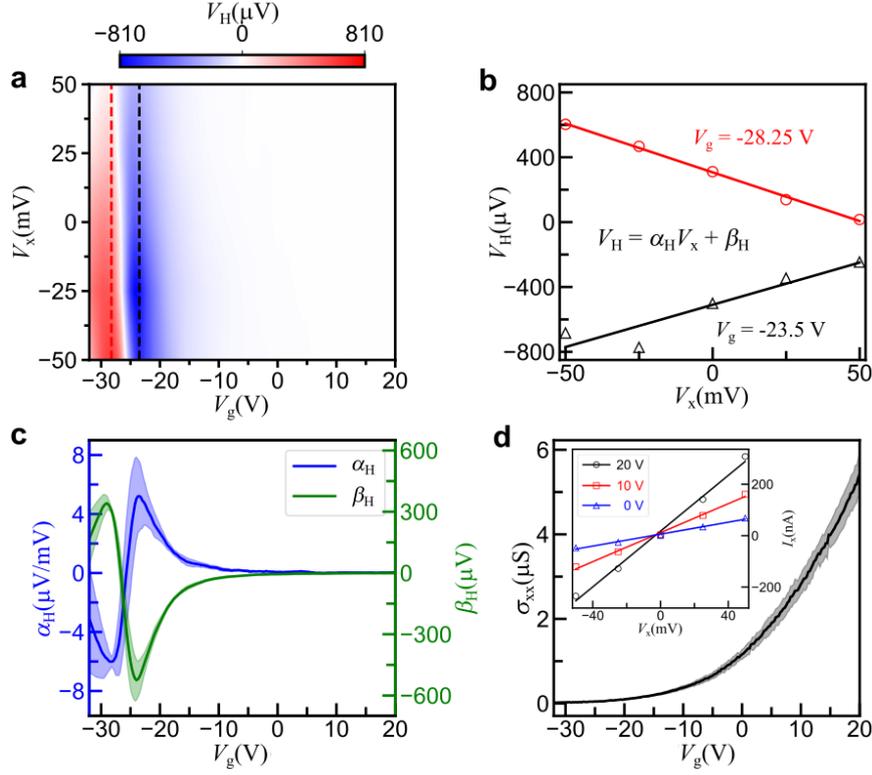

**Figure 2. Gate dependence measurement of VHE and longitudinal conductivity at 240 K. (a)** The $V_H$ as a function of $V_x$ and $V_g$. The black (red) dashed line corresponds to the black (red) symbols in (b). **(b)** The linear fitting ($V_H = \alpha_H V_x + \beta_H$) of $V_H$ versus $V_x$ at $V_g$ = -23.5 V (black line) and $V_g$ = -28.25 V (red line). **(c)** The $\alpha_H$ (proportional to VHE) and $\beta_H$ (proportional to CPC) as a function of $V_g$. The shaded area represents a 95% confidence interval. **(d)** Gate dependence of the longitudinal conductivity ($\sigma_{xx}$). The inset in (d) is the linear fitting of $I_x$ versus $V_x$ at $V_g$ = (0, 10, 20 V).

## Photocurrent gate dependence

To unveil the mechanism of VHE, we also investigate the gate dependence of the photocurrent under the unpolarized excitation. The setup of this measurement is shown in Fig.

3a. An optical chopper controlled by a second function generator is added. A bias voltage $V_y$ (between electrode 1 and 4) is applied. While the excitation is still modulated between the left and right circular polarization, the lock-in amplifier is in-sync with the chopper modulation instead of the LCR one. Hence, the measured $V_{PC}$ is the difference between the measured voltage under unpolarized optical excitation and the measured voltage under no optical excitation.

The value of $V_{PC}$ is measured as $V_g$ is swept from -32 V to 20 V at different $V_y$ values ranging from -50 mV to 50 mV (Fig. 3b). Two line-cuts in Fig. 3b at $V_g = -8$ V and $V_g = -19$ V are shown in Fig. 3c. The data are fitted with the function $V_{PC} = \alpha_{PC} V_y + \beta_{PC}$. The plots of $\beta_{PC}$ (proportional to the unpolarized photocurrent (PC)) as a function of $V_g$ are shown in Fig. 3d (see Fig. S7 in the Supplementary Information for the plot of $\alpha_{PC}$ (proportional to the photoconductivity) versus $V_g$). As can be seen from this figure, the unpolarized PC does not show polarity change confirming that the VHE polarity's gate-tunability is related to the valley optical selection rule. Moreover, by comparing Fig. 3d and Fig. 2c, we can see that **(1)** VHE vanishes when the PC is negligible, and **(2)** PC reaches maximum magnitude when VHE changes polarity. We also note that the observed gate dependence of the PC (Fig. 3d) and the longitudinal conductivity (Fig. 2d) is very similar to the one observed in Ref [30], which indicates similar PC mechanisms.

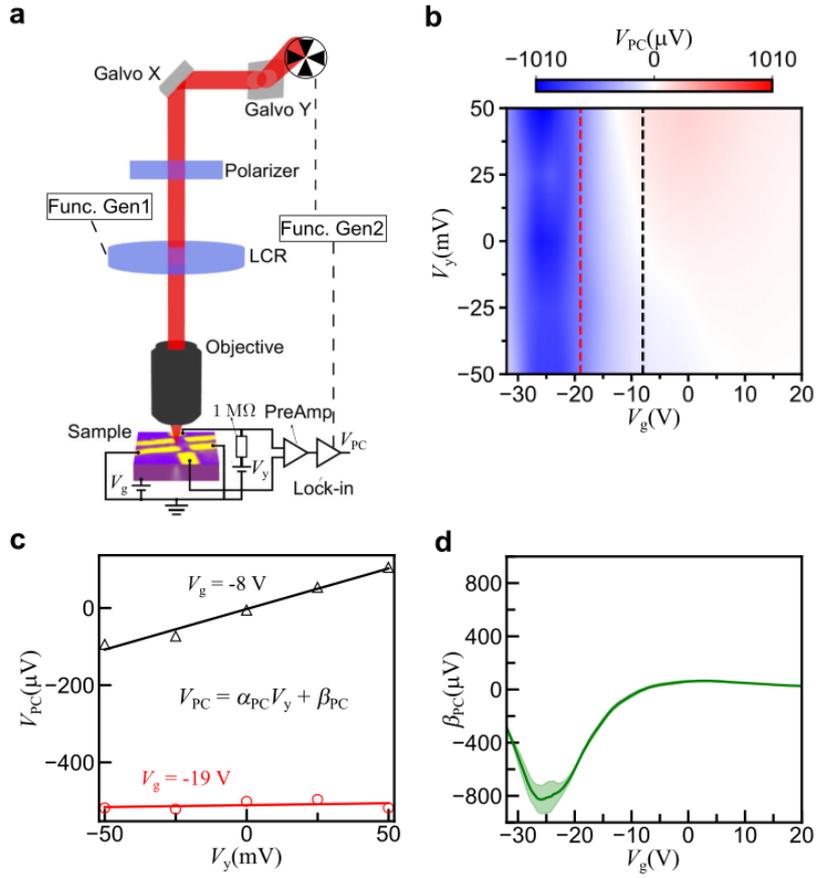

**Figure 3. Photocurrent gate dependence at 240 K. (a)** Experimental setup. **(b)** The $V_{PC}$ as a function of $V_y$ and $V_g$. The black (red) dashed line corresponds to the black (red) symbols in (c). **(c)** The linear fitting ($V_{PC} = \alpha_{PC} V_y + \beta_{PC}$) of $V_{PC}$ versus $V_y$ at $V_g = -8$ V (black line) and $V_g = -19$ V (red line). **(d)** The $\beta_{PC}$ (proportional to PC) as a function of $V_g$. The shaded area represents a 95% confidence interval.

## Bipolar nature of VHE mechanism

The gate voltage controls the electron and hole contribution to PC, as illustrated in Fig. 4(a-e). At high positive gate voltages (Fig. 4e), the photogenerated electron cannot be collected due to the in-plane lateral Schottky barrier near the electrode-MoS$_2$/WSe$_2$ interface[31], while the photogenerated hole has a short relaxation time due to the large free electron density and the occupied defect state in WSe$_2$ layer[32,33] (see Supplementary Note II for discussion on band structure and defect state energy levels[33]). This results in a negligible PC. As the gate voltage is reduced, the defect states in WSe$_2$ are emptied, and the free electron density is reduced, resulting in a longer relaxation time for the photogenerated hole and, thus, higher PC. Hence, in this regime, the PC is dominated by the contribution from the hole in $WSe_2$ (i.e., photohole-dominated regime (region II), Fig. 4d). Reducing the gate voltage further, the system enters the transition regime (shaded region, Fig. 4c), where both photogenerated electron and hole can be collected, increasing the PC. As the gate voltage is reduced more, the system enters a regime where the photogenerated hole cannot be collected due to the Schottky barrier, reducing the PC. In this regime, PC is dominated by the photogenerated electron transport contribution in the $MoS_2$ layer (i.e., photoelectron-dominated regime (region I), Fig. 4b). These four regimes (Fig. 4(b-e)) are observed in our experiment. If the gate voltage is reduced further, the p-doped regime can be reached (Fig. 4a, unobserved in the experiment). In this regime, the photogenerated electron relaxation time will be significantly reduced due to the vacant defect states in $MoS_2$ layer[32-34] and large free hole density[32], resulting in negligible PC.

The PC mechanism described above is consistent with the VHE result caused by valley-polarized electrons and holes, as illustrated in Fig. 4(f-j). At a high gate voltage, the photogenerated electron and hole cannot be collected (Fig. 4j), resulting in a negligible VHE. As the gate voltage is reduced, the system enters region II (Fig. 4i), where the contribution from the photogenerated hole dominates the VHE. Eventually, it enters the transition regime (shaded area, Fig. 4h), where both hole and electron contribute to the VHE. In contrast to the PC, which reaches its maximum magnitude in the transition regime, the VHE reaches a point where it vanishes due to the opposite VHE contribution of photogenerated electron and hole. Such opposite contributions happen when the VHE is caused by the electrically induced valley-dependent band shift[35]. As illustrated in Fig. 4 (middle right), due to the band shift, the energy band slope for photogenerated electron and hole at one valley have the same sign, which means their valley-dependent velocity components have the same direction. This mechanism agrees with the sample's CPC map (see Fig. S9 and Supplementary Note III in the Supplementary

Information). As the gate voltage is reduced further, the system enters region I (Fig. 4g), where the contribution from the photogenerated electron dominates the VHE. Our experiment stops at this regime. In a regime with even more negative gate voltage, the system will enter the p-doped regime. In this regime (Fig. 4f), the photogenerated electron and hole cannot be collected, resulting in negligible VHE. Other possible contributions to VHE are discussed in Supplementary Note III. Additional data and discussion on the temperature dependence support our interpretation (see Supplementary Note IV).

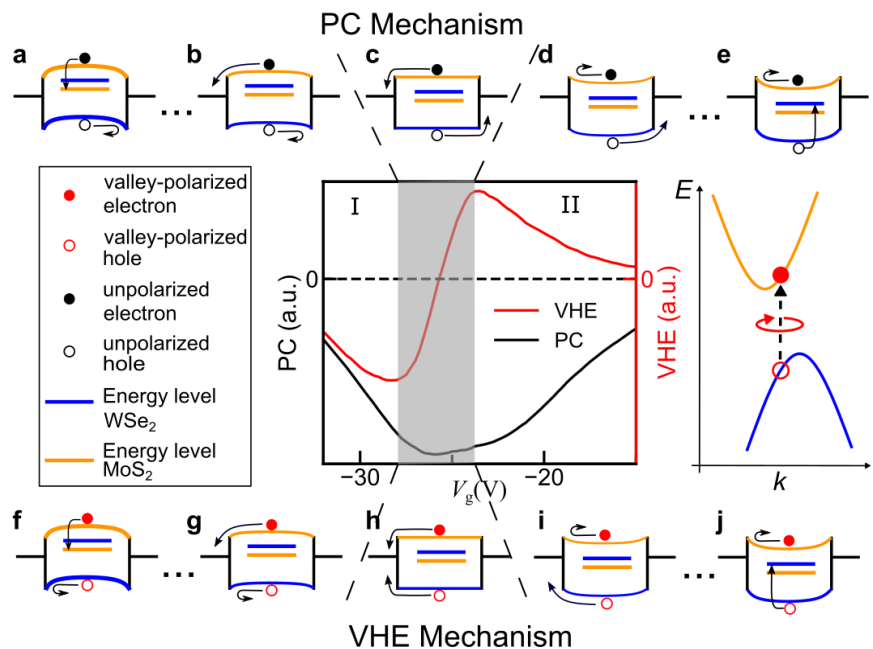

**Figure 4. VHE and PC gate dependence mechanism.** Middle left: legend. Middle center: VHE and PC as functions of the gate voltage. Middle right: due to the band shift, both the electron and hole generated by circularly polarized light have valley-dependent velocity components in the same direction. **(a-e)** PC mechanism. **(f-j)** VHE mechanism. At high negative gate voltage (a, f), photogenerated electron and hole cannot be collected, resulted in negligible VHE and PC. In Region I (b, g), the carrier transport is dominated by the photogenerated electron transport in the $MoS_2$ layer. In the transition (shaded) region (c, h), both photogenerated electron and hole contribute. The PC reaches maximum value because the valley-independent velocity component of electron and hole have opposite direction, while the VHE changes polarity because the valley-dependent velocity components of electron and hole have the same direction. In Region II (d, i), the carrier transport is dominated by the photogenerated hole transport in the WSe$_2$ layer. At high positive gate voltage (e, j), photogenerated electron and hole cannot be collected, resulted in negligible VHE and PC.

## Conclusion

In summary, we have demonstrated a robust room-temperature valley Hall effect in MoS$_2$/WSe$_2$ heterostructures. We have shown that the back-gate voltage can control both the polarity and the magnitude of the Hall voltage in our valleytronic transistor device. We attributed this gate dependence to the bipolar nature of the carrier transport and the electrically induced valley dependent band shift in the heterostructure. The full gate tunability can be used to improve the ON/OFF ratio of the valleytronic transistor and to realize more compact valleytronics logic circuits (see Supplementary Note V). Additionally, the layer- and carrier-dependent VHE demonstrated here enables the separation of the contribution from different carrier types to the valley current, which adds to the toolbox for carrier transport characterization[36]. Moreover, recently, excellent charge transport and the possibility of flexible stacking between metal and 2D materials have been achieved[37,38], and a designated substrate can facilitate the valley transport[39,40]. Together with these facts, our results can provide novel platforms for valleytronic devices and carrier characterization in 2D material in the future.

## Methods

<u>Sample fabrication</u>

The MoS$_2$/WSe$_2$ heterobilayer is fabricated using the standard dry transfer method. Monolayers are firstly exfoliated on polydimethylsiloxane (PDMS) films. The PDMS films with monolayer are then sequentially aligned and stamped on a silicon substrate with a 300 nm SiO$_2$ overlayer to make a van der Waals heterobilayer. The metal electrodes, consisting of 5 nm thick Cr and

50 nm thick Au, are fabricated using standard electron beam lithography (EBL) and thermal evaporation techniques.

Optical setup

The heterobilayer sample is mounted in a closed-cycle cryostat for room temperature and low-temperature experiments. A tunable Ti-sapphire laser is used as the excitation source. The laser is directed by a single-mode fiber to the homemade setup. A galvanometer scanner is used for scanning the laser beam over the heterobilayer region. A polarizer and a liquid crystal retarder are mounted after the scanner to perform polarization modulation. The laser beam is focused by an objective to sub-µm in diameter on the sample surface. Since the heterobilayer region is small (~2 µm × 6 µm), the laser beam scanning over this region can be approximated as a normal incidence excitation.

Electrical measurement setup

The longitudinal voltage is applied using a Keithley 2636B source meter. The Hall signal and photocurrent signal are amplified by a differential preamplifier with an amplification factor of 500 (A-M Systems, Inc. Model 3000) and a lock-in amplifier with an amplification factor of 1000. The lock-in amplifier references the frequency of the liquid crystal retarder. A square wave of 33 Hz is used to modulate the liquid crystal retarder for polarization modulation.

**Acknowledgments** We thank the discussion with Feng Wang, Wang Yao. acknowledge the financial support from the Singapore National Research Foundation through its Competitive Research Program (CRP Award No. NRF-CRP21-2018-0007, NRF-CRP22-2019-0004), Singapore Ministry of Education (MOE2016-T2-2-077, MOE2016-T2-1-163, MOE2016-T3-

1-006 (S)). Q.X. gratefully acknowledges National Natural Science Foundation of China (No. 12020101003) and start-up grant from Tsinghua University.

# Supplementary information

## Supplementary Note I. Determination of MoS$_2$/WSe$_2$ samples stacking alignment

The stacking sequence can be determined from the second harmonic generation measurement (SHG). For the AA stacking, the SHG from the heterostructure is stronger than the SHG from the individual monolayers, while for AB stacking, it is weaker[1]. The SHG measurement from Sample 1 and Sample 2 are shown in Fig. S1(a, b). From these figures, it can be seen that the SHG from the heterostructure is stronger than from the individual monolayer, indicating that our samples are AA-stacked heterostructures.

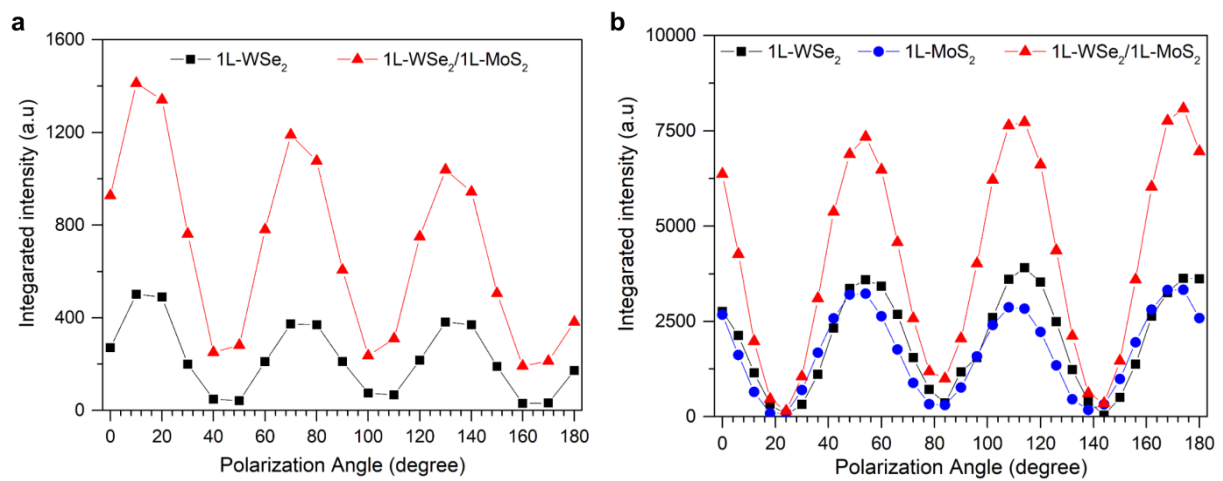

**Figure S1. SHG measurement of MoS$_2$/WSe$_2$ samples. (a)** Sample 1. The SHG measurement results for the monolayer MoS$_2$ region and heterostructure region are shown by the black and red symbols, respectively. **(b)** Sample 2. The SHG measurement results for the monolayer WSe$_2$ region, monolayer MoS$_2$ region, and heterostructure region are shown by the black, blue, and red symbols, respectively. In both Sample 1 and Sample 2, the heterostructure's SHG signal is stronger than those from the monolayers, indicating AA-stacking alignment.

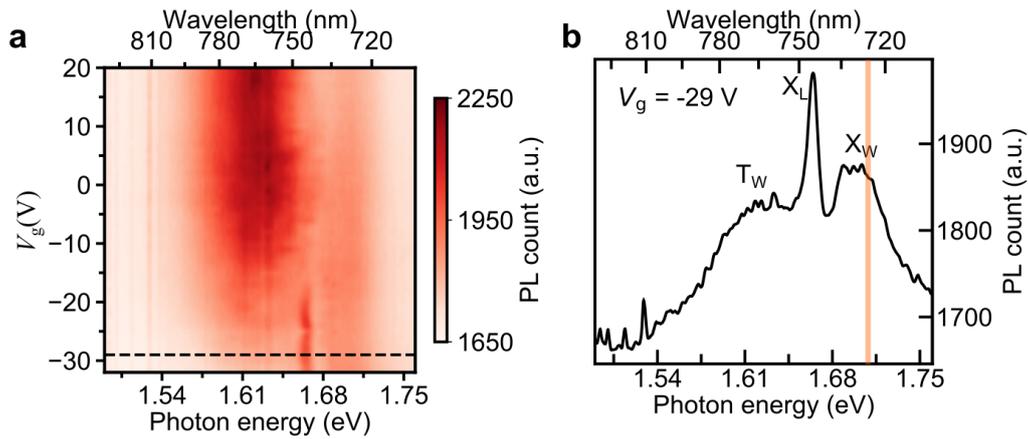

**Figure S2. Photoluminescence spectrum of MoS$_2$/WSe$_2$ heterostructure at 140 K. (a)** PL spectrum as a function of the gate voltage. The black dashed line cut corresponds to the spectrum shown in (b). **(b)** PL spectrum at $V_g = -29$ V. The neutral, local, and charged exciton are marked as X$_W$, X$_L$, and T$_W$, respectively. The orange line indicates the 726 nm excitation laser used in the main text.

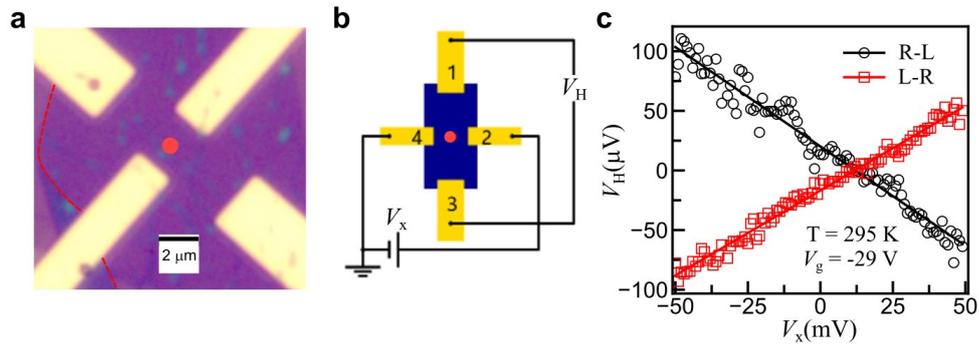

**Figure S3. Room temperature VHE in another MoS$_2$/WSe$_2$ sample (Sample 2). (a)** Optical image of the sample. The red dot is the 220 μW, 726 nm optical excitation location, while the red dashed line is the heterostructure sample boundary. **(b)** Electronic configuration for Hall voltage measurement. The red dot is the excitation location **(c)** Observation of VHE at room temperature. The black and red symbols are the measured difference between the Hall voltage under the right and left circular polarization (R-L) and under the left and right circular polarization (L-R), respectively. The lines are linear fitting.

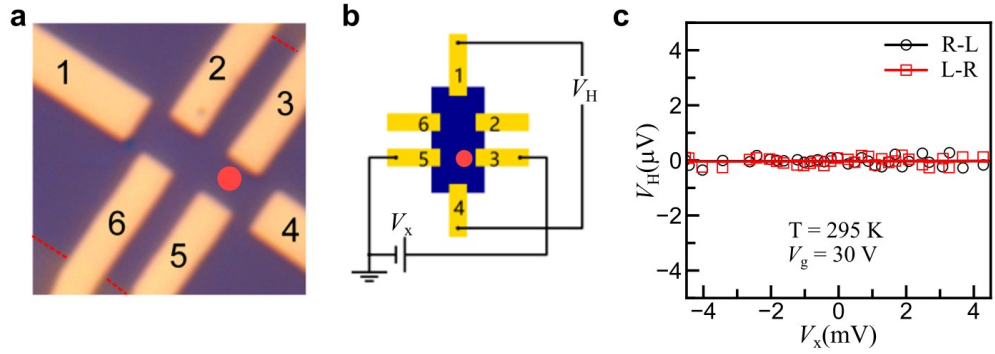

**Figure S4. Negligible room temperature VHE in the MoS$_2$ monolayer sample.** **(a)** Back-reflection image of the sample. The red dot is the 11 μW, 726 nm optical excitation location, while the red dashed line is the heterostructure sample boundary. **(b)** Electronic configuration for Hall voltage measurement. **(c)** Negligible VHE at room temperature. The black and red symbols are the measured difference between the Hall voltage under the right and left circular polarization (R-L) and under the left and right circular polarization (L-R), respectively. The lines are linear fitting. A gate voltage of 30 V is applied to increase the magnitude of VHE (see Fig. S5 for the gate dependence of VHE in the MoS$_2$ monolayer sample at 140K).

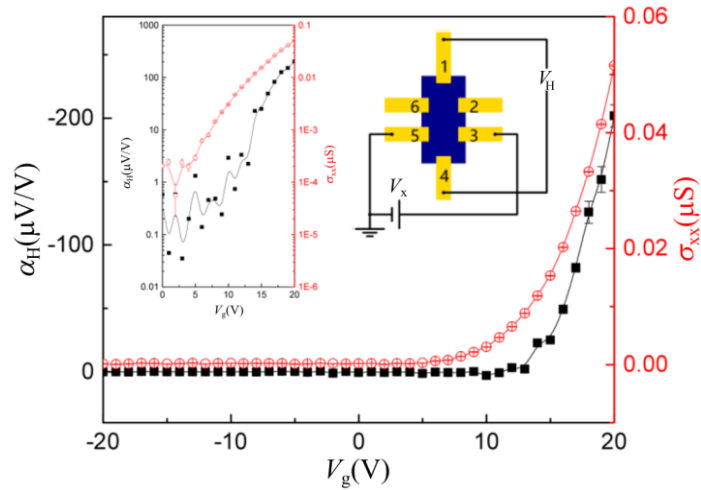

**Figure S5. Gate dependence of VHE and longitudinal conductivity in the MoS$_2$ monolayer sample at 140K.** VHE (black symbols and line) and longitudinal conductivity (red symbols and line) versus the back-gate voltage. The inset shows a semi-log plot of the same back-gate voltage dependence. Only VHE magnitude is gate tunable.

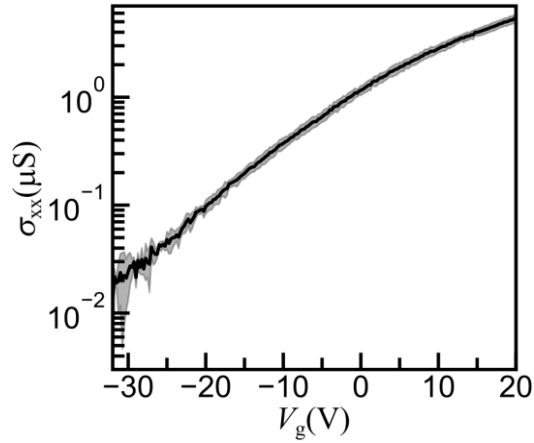

**Figure S6. Log-linear plot of Sample 1 longitudinal conductivity gate dependence at 240 K.** Within the considered gate voltage, the longitudinal conductivity is > 0.01 µS, corresponding to sample resistance of < 100 MΩ.

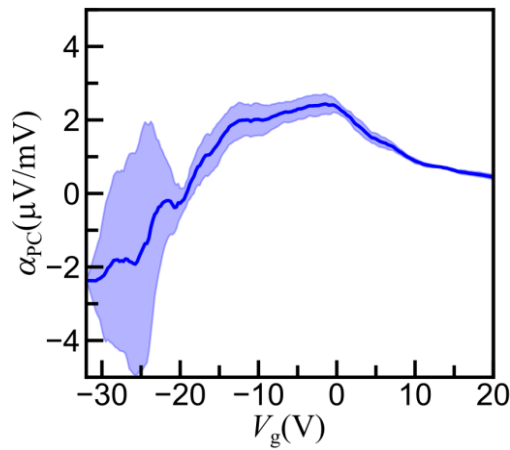

**Figure S7. Photoconductivity gate dependence at 240 K.** The $\alpha_{PC}$ (proportional to PC) obtained from the linear fitting ($V_{PC} = \alpha_{PC} V_y + \beta_{PC}$) is plotted as a function of $V_g$. The shaded area represents a 95% confidence interval.

## Supplementary Note II. Band structure and defect state energy levels

In Fig. S8, the band structure of the MoS$_2$/WSe$_2$, including the energy level range for the defects states in MoS$_2$ ($e$ and $e$-$e$ defect state from sulfur-atom vacancy and divacancy) and WSe$_2$ ($e$ defect state from selenium-atom vacancy)[2], is shown together with the work function of the metal electrode (Cr and Au)[3]. Here, we use 4.2 eV and 3.6 eV as the electron affinity in monolayer MoS$_2$ and WSe$_2$, respectively[4,5]. The energy level values are their value in eV with respect to the vacuum energy level. The orange, blue, and grey (black) represents the energy levels in MoS$_2$, WSe$_2$, and electrode, respectively. The shaded areas are the possible range of energy levels for the defects and the electrode's effective Fermi level, while the solid lines are their most likely energies.

For WSe$_2$, the defect state is split due to the spin-orbit coupling[2]. Regarding the electrode's effective Fermi level (solid black line), we take it to be the average of the Cr and Au Fermi levels. The solid lines are used in Fig. 4 in the main text (note: only one WSe$_2$ defect level is shown in that figure for simplicity). The dashed purple line is the intrinsic Fermi level of the heterostructure.

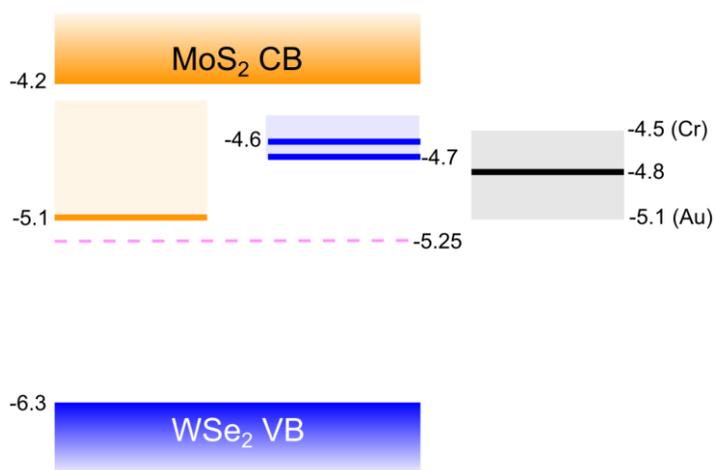

**Figure S8. MoS$_2$/WSe$_2$ band structure and defect state energy levels.** The energy level values are their value in eV with respect to the vacuum energy level. The orange, blue, and grey (black) represents the energy levels in MoS$_2$, WSe$_2$, and electrode, respectively. The shaded areas are the possible range of energy levels for the defects and the electrode's effective Fermi level, while the solid lines are their most likely energies.

**Supplementary Note III: More discussion on CPC map and optoelectronic response**

From the fact that our sample has AA stacking alignment (see Fig. S1 and Supplementary Note I), we argue that the primary driving mechanism behind the VHE is not the conventional intrinsic anomalous velocity[6], which can be expressed as $\vec{v}_A^{conv} = \pm e/\hbar \, \vec{E} \times \vec{\Omega}_{n,\vec{k}}$ with $e$ is the electron charge, the + (-) sign is for the hole (electron), $\hbar$ is the reduced Planck constant, $\vec{E}$ is the electric field, $\vec{\Omega}_{n,\vec{k}}$ is the Berry curvature at band $n$ and momentum $\vec{k}$. Unlike the AB-stacked case[7,8], in an AA-stacked $MoS_2/WSe_2$, the Berry curvatures of the conduction band in $MoS_2$ and valence band in $WSe_2$, at the same momentum, have the same sign. As a result, the valley electron's anomalous velocity in the $MoS_2$ layer has an opposite direction to the valley hole's anomalous velocity in the $WSe_2$ layer. This argument contradicts the experimental results that indicate that the valley-dependent velocity of electron and hole must have opposite direction. Therefore, our results mean that the primary mechanism behind VHE in our sample is not the conventional anomalous velocity.

To know the VHE and CPC mechanism in more detail, we obtain the low temperature (140 K) normalized CPC map for Sample 1, as shown in Fig. S9a. It is obtained using a 7 μW, 726 nm optical excitation. No back gating is applied. The CPC is collected using electrode 1 and 4 with positive CPC means that current is flowing into electrode 1 from the sample. The simulation result of the CPC map using the theoretical model described in Ref [9,10] (Fig. S9b) shows a good agreement with the experimental result.

In this model, the in-plane electric field causes the valence band to shift with the shift's direction is opposite for K and K' valleys. Using the normal velocity relation[11] $\vec{v}_c = \nabla_{\vec{k}} \varepsilon_n(\vec{k}) / \hbar$, where $\vec{v}_c$ is the carrier (electron or hole) velocity and $\varepsilon_n(\vec{k})$ is the electron energy at band $n$ and momentum $\vec{k}$, it can be shown that the band shift-induced velocity is proportional to the shift. The shift always has a component perpendicular to the electric field resulted in the band shift-induced anomalous velocity, while a strain influence can cause an additional component parallel to the electric field. The excellent agreement between the theoretical and experimental results in Fig. S9 indicates that the CPC is caused by this electrically induced valence band shift. Based on Fig. 2c in the main text, the VHE and CPC have a similar origin. Hence, the VHE should also be caused by the electrically induced valence band shift.

The CPC map can also be used to analyze the trion contribution. Based on the map, the CPC is strongest near the edge of the electrode. As has been shown in Fig. 2c in the main text, VHE is proportional to CPC. A strong VHE near the edge shows that the VHE is caused by free carriers resulted from the exciton dissociation instead of trions[12]. Hence, we can neglect the trion contribution for VHE under 726 nm optical excitation.

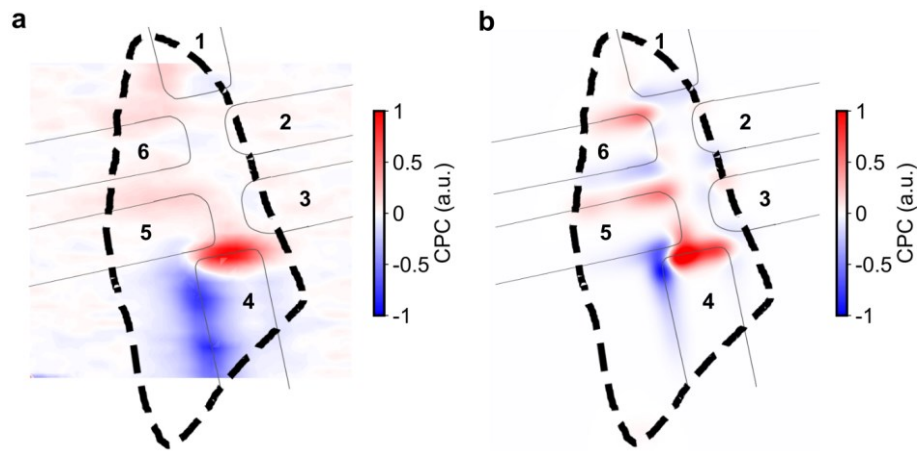

**Figure S9. CPC map at 140 K for Sample 1. (a)** Experimental result. The CPC map shows normalized CPC at 140 K with the electronic configuration as in Fig. 1c in the main text. A 7 µW, 726 nm optical excitation is used. No back gating is applied. The electrodes' edges and the heterostructure region boundary are illustrated by solid and dashed black lines, respectively. The electrode numbering is the same as in Fig. 1b in the main text. Electrode 1 and 4 are used to collect the CPC. **(b)** Simulation result.

Similar to the case in Ref [9], the nonzero PC and CPC under normal incidence signify that our system's symmetry has, at most, only one mirror symmetry, and the $C_3$ symmetry of TMD is broken[13-15]. This reduced symmetry can be understood from three facts: **(1)** the TMD heterostructure can have incommensurate stacking resulted in the absence of any rotation and mirror symmetry[16], **(2)** the symmetry breaking can be induced by the strain in the sample, and **(3)** the excitation location is not symmetrical with respect to the two current-collecting electrodes (electrode 1 and 4).

Lastly, we note that the mechanism in Fig. 4 in the main text does not apply to the longitudinal conductivity and the unpolarized photoconductivity. This mechanism only works when the Schottky barrier's blocking effect is dominant. On the contrary, the linear behavior of the longitudinal transfer characteristics (i.e., $I_x$ versus $V_x$, inset of Fig. 2d in the main text) indicates that the Schottky barrier blocking effect is negligible for the longitudinal transport driven by an external voltage source[17]. Additionally, the mechanism in the main text's Fig. 4 also does not apply when the excitation location is in the Schottky barrier regime (depletion regime). At this location, the electric field due to the Schottky barrier will contribute to the photo-induced emf negating the barrier's blocking effect. This reasoning indicates that the gate dependence depends on the excitation location, which should be investigated in further studies.

**Supplementary Note IV. Gate dependence at low temperature (140 K)**

From the VHE and PC mechanism described in the main text, we can predict that both the magnitude and polarity of the VHE at low temperature should also be gate tunable. Moreover, VHE and CPC at low temperature should also be correlated. However, while at high temperature, the hole mobility in WSe$_2$ can be comparable to the electron mobility in MoS$_2$, at low temperature, the hole mobility is much smaller than electron mobility[18,19]. Consequently, the magnitude of the extremum points (maximum positive or negative) of the VHE and CPC will be largely different from each other. Since electron mobility at low temperature is much larger than the hole's, the extremum points of VHE and CPC at photoelectron dominated regime (more negative gate voltage) will also be much larger than the one at photohole dominated regime (more positive gate voltage). This difference in carrier mobility will also mean that the photohole contribution to the PC is negligible. As a result, the peak of the PC should happen at the same gate voltage as the VHE extremum point at the photoelectron dominated regime.

The experimental result regarding the gate dependence of VHE, CPC, and PC at 140 K are shown in Fig. S10(a, b). The excitation location is as in Fig. 1d in the main text. As in the main text, $V_H = \alpha_H V_x + \beta_H$, and $V_{PC} = \alpha_{PC} V_y + \beta_{PC}$. Hence, VHE, CPC, and PC are proportional to $\alpha_H$, $\beta_H$, and $\beta_{PC}$, respectively. From Fig. S10a, we can see that the VHE magnitude and polarity's gate tunability is also observed at low temperature. The VHE and CPC are zero and optimum (i.e., maximum or minimum) at the same gate voltage, showing a strong correlation between them. We can also see that the magnitude of the peak at $V_g \sim$ -25 V is much larger than the dip at $V_g \sim$ -14 V. From Fig. S10b, we can see that the peak of the PC also happens at $V_g \sim$ -25 V. These observations are in line with our prediction, supporting the VHE and PC mechanism described in the main text.

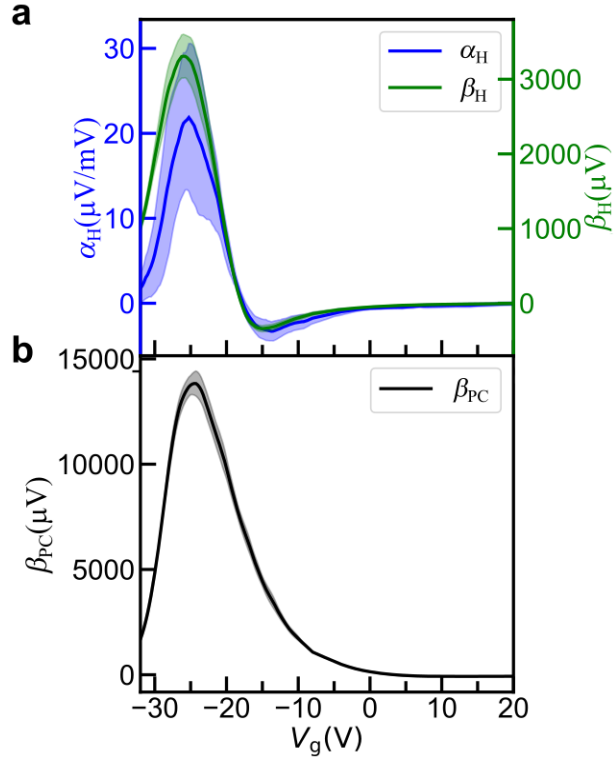

**Figure S10. Gate dependence of VHE, CPC, and PC at 140 K for Sample 1. (a)** The $\alpha_H$ (proportional to VHE) and $\beta_H$ (proportional to CPC) as a function of $V_g$. The shaded area represents a 95% confidence interval.

**(b)** The $\beta_{PC}$ (proportional to PC) as a function of $V_g$. The shaded area represents a 95% confidence interval.

## Supplementary Note V. Application of full gate tunability

In this section, we discuss potential applications of the full gate tunability (i.e., both the polarity and magnitude are gate tunable). Here we consider the doping control by an offset gate voltage.

**(1)** A valleytronic transistor with no limitation on the ON/OFF ratio

The valleytronic transistor operation is depicted in Fig. S11a. In a valleytronic transistor, the bias voltage $V_x$ remains constant while the gate voltage $V_g$ acts as an input and the Hall voltage $V_H$ is the output. A constant gate voltage offset $V_{g0}$ can be used to control the initial doping. A positive Hall voltage indicates an ON state, while a zero Hall voltage indicates an OFF state. In the MoS2 valleytronic transistor, the ON/OFF ratio is limited by the nonzero minimum Hall voltage (Fig. S11b). On the contrary, in the MoS$_2$/WSe$_2$ valleytronic transistor, since the Hall voltage's polarity can be changed from positive to negative, zero Hall voltage can be achieved (Fig. S11c). Consequently, there is no limitation on the ON/OFF ratio.

**(2)** A more compact two-input valleytronics logic circuit

The Hall bar structure can also be used as a two-input valleytronics logic gate. In this case, both the bias voltage $V_x$ and gate voltage $V_g$ act as inputs while Hall voltage $V_H$ is the output (see Fig. S12a). The bias voltage is kept at zero when the logic gate is unused to reduce power consumption. A constant gate voltage offset $V_{g0}$ is used to control the initial doping.

Due to the full gate tunability, the MoS$_2$/WSe$_2$ valleytronics logic gate is more versatile than the MoS$_2$ one. This can be seen by comparing the possible operation regimes and truth tables of these two logic gates (see Fig. S11(b-d)). Here, a negative voltage represents '1', while a zero or positive voltage represents output '0'. The MoS$_2$ Hall bar structure can only be used as a NAND gate (Fig. S11b), while the MoS$_2$/WSe$_2$ can be used as a NAND (Fig. S11c) or XNOR gate (Fig. S11d) depending on the offset gate voltage value. Since an XNOR gate can only be created using at least five NAND gates, a MoS$_2$/WSe$_2$ valleytronic logic circuit will have fewer components than the MoS$_2$ one.

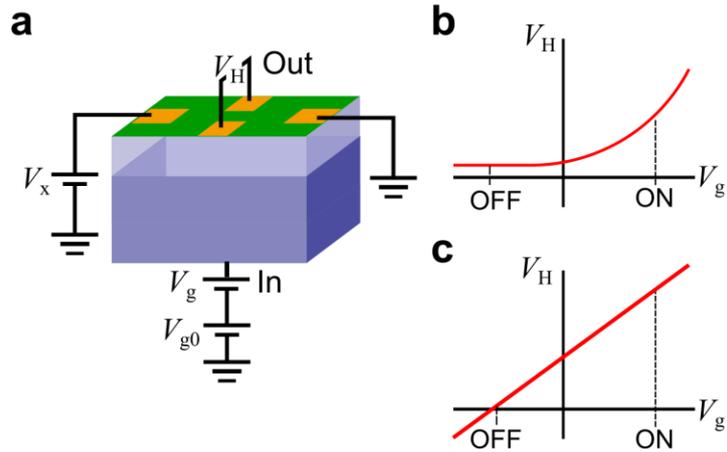

**Figure S11. Valleytronic transistor.** (a) Valleytronic transistor schematic. The bias voltage $V_x$ and the offset gate voltage $V_{g0}$ remains constant. The input is the gate voltage $V_g$ and the output is the Hall voltage $V_H$. (b) MoS$_2$ valleytronic transistor transfer characteristic. The minimum value of $V_H$ for the OFF state is finite, limiting the ON/OFF ratio. (c) MoS$_2$/WSe$_2$ valleytronic transistor transfer characteristic. The minimum value of $V_H$ for the OFF state can reach zero, removing the limitation on the ON/OFF ratio.

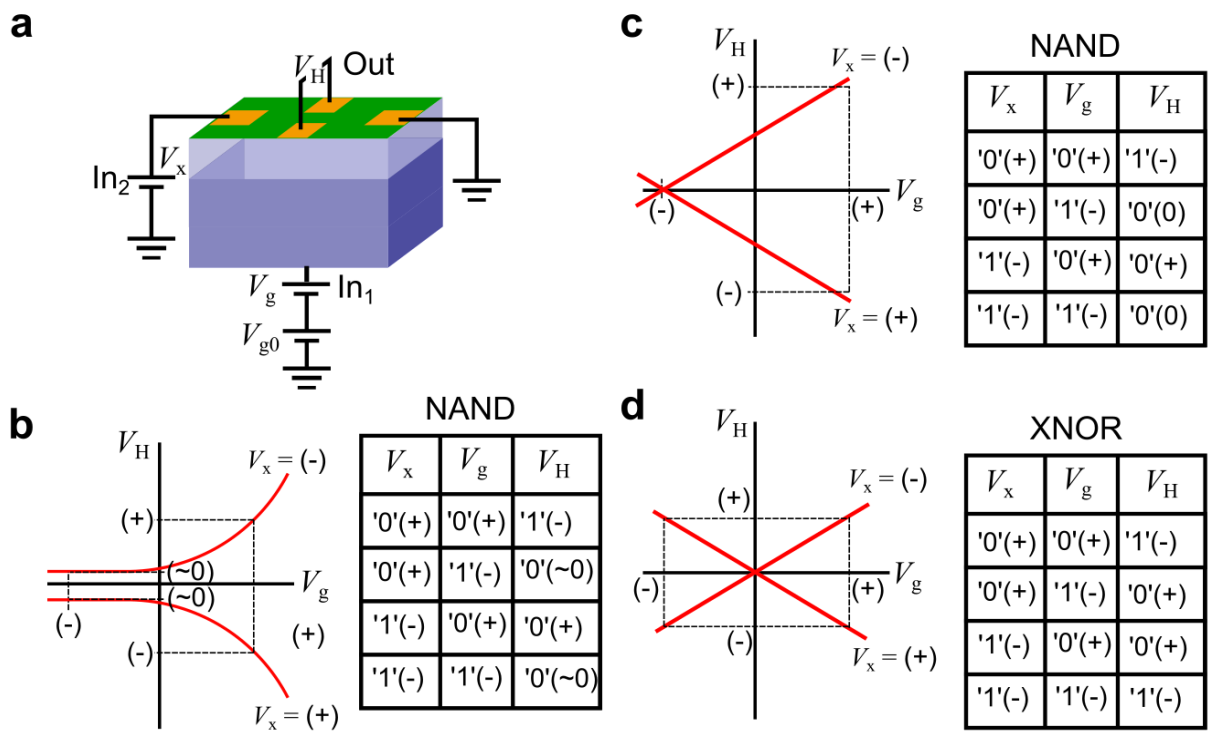

**Figure S12. Valleytronic logic gate. (a)** Valleytronic logic gate schematic. The offset gate voltage $V_{g0}$ remains constant. The inputs are the gate voltage $V_g$ and the bias voltage $V_x$. The output is the Hall voltage $V_H$. **(b)** MoS$_2$ NAND valleytronic logic gate. The illustration of $V_H$ versus $V_g$ at two different $V_x$ are shown. **(c)** MoS$_2$/WSe$_2$ NAND valleytronic logic gate. **(d)** MoS$_2$/WSe$_2$ XNOR valleytronic logic gate. The value of $V_{g0}$ here is different from the one in (c).